\documentclass[aps,prx,twocolumn,floats,showpacs,superscriptaddress,nofootinbib]{revtex4-1}
\usepackage{graphicx,epsfig}
\usepackage{times,bbm}
\usepackage{graphics,dcolumn,bm,float}
\usepackage{amssymb,amsmath,rotate,color,amsfonts}
\usepackage[title,titletoc,toc]{appendix}
\usepackage{mathtools}
\usepackage{booktabs}
\usepackage{tcolorbox}
\usepackage[pagebackref=false,colorlinks,linkcolor=magenta,citecolor=blue,urlcolor=magenta]{hyperref}

\usepackage{wrapfig}
\usepackage{lipsum}
\usepackage{mwe}
\usepackage[mathlines]{lineno}
\usepackage[mathscr]{euscript}
\usepackage{hyperref}
\usepackage{breqn}
\usepackage{bbold}
\usepackage{pgfplots}
\usepackage{float}
\usepackage{tkz-euclide}
\usepackage{braket}
\usepackage{physics}
\usepackage[caption=false]{subfig}
\usepackage[export]{adjustbox}
\usepackage{tikz}
\usepackage{slashed}
\usepackage{bm}
\usepackage{multirow}
\usetikzlibrary{through,calc}
\usetikzlibrary{positioning}

\newcommand{\be}{\begin{equation}}
\newcommand{\ee}{\end{equation}}

\newcommand{\bearr}{\begin{eqnarray}}
\newcommand{\eearr}{\end{eqnarray}}
\newcommand{\nn}{\nonumber}

\newcommand{\bs}{\boldsymbol}
\newcommand{\bmt}{\left[\begin{matrix}}
\newcommand{\emt}{\end{matrix}\right]}

\begin{document}

\title{Semiclassical transport in two-dimensional Dirac materials with spatially variable tilt}
\author{Abolfath Hosseinzadeh}
\affiliation{Department of Physics$,$ Faculty of Basic Sciences$,$ Shahed University$,$ Tehran 18151/159$,$ Iran}
\author{S. A. Jafari}%
 \email{akbar.jafari@rwth-aachen.de }
\affiliation{Peter Gr\"unberg Institute 2$,$ Forschungszentrum J\"ulich$,$ 52425 J\"ulich$,$ Germany}
\affiliation{2nd Institute of Physics C$,$ RWTH Aachen University$,$ 52074$,$ Aaachen$,$ Germany}




\date{\today}

\begin{abstract}
We use Boltzmann theory to study the semi-classical dynamics of electrons in a two-dimensional (2D) tilted Dirac material in which the tilt varies in space. 
The spatial variation of the tilt parameter induces a non-trivial spacetime geometry on the background of which the electrons roam about. As the first manifestation of 
graivto-electric phenomena, we find a geometric planar Hall effect according to which a current flows in a direction transverse to the chemical potential gradient and 
is proportional to $g^{xy}$ component of the emergent spacetime structure. The longitudinal conductivity contains information about the gravitational red-shift factors. 
Furthermore, in the absence of externally applied electric field there can be "free-fall" or zero-bias currents that can be used as detectors of terahertz radiation.
\end{abstract}

\maketitle


\section{\label{sec:level1}Introduction}
Before the emergence of Dirac materials~\cite{wehling2014dirac,katsnelson2012graphene}, the Dirac equation was thought to 
describe behavior of elementary particles at extremely high energies and velocities comparable to 
the speed of light where relativistic effects become inevitable. Dirac solids differ from other solids in that they offer
"Dirac fermions" at mundane sub-eV energy scales~\cite{Fukuyama2012,Isobe2022,Fransson2016}. Interestingly their sub-eV
energy scale makes them attractive for sub-eV dark matter detection~\cite{Geilhufe2020}.

What else can a Dirac solid offer that the rest of solids can not provide? To answer this question, let us note that
the typical dispersion relation $\hbar^2k^2/(2m^*)$ of ordinary conductors versus the dispersion $v_F\sqrt{|\bs k|^2+(mv_F)^2}$ of Dirac solids
is reminiscent of the difference between Galilean spacetime versus the Minkowski spacetime. As such, the Dirac equation in such solids naturally
follows if one assumes an underlying Minkowski spacetime structure where the upper limit of speeds instead of $c$ is 
$v_F$~\cite{Volovik2016,Lukose2007,SaharPolarization,Gallerati2022,HornerThesis2022}. 
The nice thing about Dirac equation in solids is that starting from an emergent Lorentz-invariant formulation, one can
easily envisage violation of the Lorentz invariance~\cite{Babak2022}. In fact 
the two branches of the Dirac dispersion relation in solids correspond to upper (conduction) and lower (valence) bands of pertinent solids. As such,
it is quite natural to expect an arbitrary crossing of the two bands that deviates from being
symmetric along the energy axis~\cite{Hirata2016,Kajita2014,Kiswandhi2021}. In this way a tilted Dirac cone band crossing can be obtained. 
Even in this case one can find a generalized chiral symmetry to protect the the crossing from being gapped out~\cite{Aoki2011}. 
One way to tilt a Dirac cone in two space dimensions is to break the rotational symmetry of the continuum limit down to $C_{2v}$~\cite{Yekta2023}. 
When the lowering of symmetry to $C_{2v}$ is caused by an external effect, it can be used as a convenient tool to control the tilt of the Dirac cone. 
A very nice example of such externally induced tilting is given by magnetization induced tilting of the Dirac cone on the
surface of topological insulators~\cite{Ogawa2016} where an in-plane magnetic influence couples to the Dirac cone in a way to 
control its tilt~\cite{Jafari2023}. In this case the magnetization can be considered as the agent that externally imposes $C_{2v}$ symmetry 
on Dirac electrons. In this setting, a spatially variable magnetization of appropriate form can generate spatially variable tilt 
profile~\cite{Jafari2023}. Given that the tilt parameters specify an emergent spacetime metric~\cite{Volovik2016,SaharPolarization,Jafari2019,Kedem2020,Jafari2023}, 
a spatially variable tilt will imprint a curved spacetime where the properties of spacetime are given by the shape of variations of the tilt in the space~\cite{farajollahpour2020synthetic}. 
 
In three space dimensions (3D) Weyl semimetals the cones can be naturally tilted: The 3D Dirac solids have both time-reversal (TR) and inversion symmetries. 
Breaking either of these results in Weyl semimetals where the two chiralities get separated in the momentum space~\cite{Armitage2018}. 
Since the Lorentz invariance is not stringent in solids, the Weyl nodes can be tilted to give rise to the so called type-II Weyl semimetals~\cite{Soluyanov2015,Koepernik2016,Wang2016,Xu2017,Yao2019}. The difference between type-I and type-II Weyl semimetals can be experimentally probed~\cite{Jiang2017,Zheng2018,Ma2019}. 
The violation of Lorentz invariance by tilting of the Dirac/Weyl cone still does leave a very rich symmetry behind.
It turns out that the Minkowski spacetime is destroyed, but a much more interesting spacetime which is a deformation of
the Minkowski spacetime emerges that is determined by the amount of tilting~\cite{Volovik2016,SaharPolarization,Jafari2019,Kedem2020}.
However, in 3D it is difficult to tune the tilting by external means. That is why we focus on 2D tilted Dirac cones. 

Viewing the tilt in this way has far reaching consequences~\cite{Landau1980}. 
The fact that "tilt" deformation of the Dirac dispersion can be encoded as a deformation of the Minkowski spacetime
in a more precise language can be attributed to "frame field" (vierbein). Therefore tilting of the Dirac cone is a solid-state way of attaching vierbein fields 
to Dirac fermions~\cite{Yepez2011}. Hence the ability to design spatially non-uniform tilt will
be tantamount to designing frame fields that are not constant in space or spacetime -- similar to basis vectors on the sphere that vary from point to point --. 
In this way one obtaines a (curved) spacetime geometry for the electrons moving in such materials. As such, the 
curvature of the synthetic spacetime arises from the electrical forces and peculiar arrangements of atoms and orbitals. One can envisage that
this can provide much stronger curvature effects than the actual (weak) gravitational fields. 
Given that the entire band structure is a result of Coulomb interaction between the electron and ions, the existence of frame fields in this
context is quite non-trivial and can be considered as a paradigm for the geometric formulation of electrical forces. In fact the other side of this
analogy is known as gravitomagnetic effects~\cite{ryder2009introduction}, according to which certain aspects of gravitational forces formally resemble the electromagnetic forces.
But what concerns us in the case of tilted Dirac/Weyl cone materials is the fact (at least some parts of) the Coulomb forces may look like "gravitational" forces in the sense that
they can be encoded into a spacetime geometry (metric). The availability of such a (synthetic) curved spacetime in quantum materials~\cite{Tohid2019,farajollahpour2020synthetic} 
with tilted Dirac cone, and the promising possibilities to control the location~\cite{Nilforoushan2021} of the Dirac node and its tilting~\cite{Yekta2023} by atomic
manipulations, invites us to revisit basic solid state phenomena in an arbitrary spacetime background.

A very well formulated problems in a curved background spacetime is the study of the classical path of particles or 
geodesics~\cite{Landau1980,ryder2009introduction}. 
At a quantum level, although the study of wave propagation in curved spacetimes is a highly non-trivial problem~\cite{MASHHOON1987,Ching1995}, 
the null geodesics (classical trajectory of massless particles) provides information about the wave function~\cite{Matthew2022}. 
Within a semi-classical treatment where the energy band structure of electron arises from quantum mechanical treatment of the band theory,
while the motion of electron is governed by classical equations of motion, the form of the motion of electrons (i.e. their geodesic)
is expected to leave its fignerprint in the conductivity of these solid. 
The purpose of this paper is to formulate the conductivity of a 2D tilted Dirac material whose tilt depends on spatial coordinates
in two space dimensions. 

A note on the comparison of the Berry curvature and the curvature arising from the emergent spacetime metric is in order:
The Berry curvature arises from attaching a Bloch fiber to every point in the Brillouin zone taken to be the base manifold. In this way
one forms the Bloch fiber-bundle structure where the curvature effects arises from the twist of the wave function on the Bloch fiber. 
In the tilted Dirac cone materials with spatially variable tilt, the curvature effects arises from attaching a tangent space as 
a fiber to every point in the real space. The connections in the former case are Berry connections (or gauge fields), whereas in the
latter case they could be Christoffel connections (for bosons) or spin connections or spin gauge fields for fermions~\cite{farajollahpour2020synthetic}. 
The Berry phase effects arising from the tilted Dirac band crossing become manifest when one considers the semiclassical
theory in a background magnetic field~\cite{Young2013}. Since in this work we are only interested in the response of the electrons
to an applied electric field, we do not consider the Berry curvature and focus only on the curvature effects arising from the spatial variation
of the tilt of the Dirac cone. The situation where both of the connections are non-trivial, namely a topologically non-trivial band 
on an emergent solid-state spacetime can also be formulated which is beyond the scope of the current work. 

Inspired by recent progress in gigantic tilting of the spin-orbit coupled Dirac cones~\cite{Ogawa2016,Jafari2023}, 
our model will consists in a single tilted Dirac cone in 2D with arbitrary spatial dependence of the tilt parameter $\bs\zeta(\bs r)$
that corresponds to the metric~\cite{Martel2001,farajollahpour2020synthetic},
\begin{equation}
g_{\mu\nu}=
\begin{pmatrix}
\zeta^{2}-1 & -\zeta_{x} & -\zeta_{y}\\
-\zeta_{x} & 1 & 0\\
-\zeta_{y} & 0 & 1
\end{pmatrix}.
\label{Metric}
\end{equation} 
The ability to synthesize a spacetime metric with a given $\bs\zeta(\bs r)$ is tantamount to implementing certain built-in background electric fields
that are encoded  in the Christoffel connections of the above metric. Such fields are physically similar to the built-in electric fields that arise in
pn-junctions~\cite{girvin2019modern,Shockley1949}. But since in the present case such built-in electric fields are acting on Dirac electrons,
their effect can be nicely described with appropriate Christoffel connections. 
As such in a quantum material with a spatially variable tilted Dirac cone,
the background electric fields that are encoded into the spacetime metric (ultimately arising from ions) guide the electrons in certain paths even if 
no probe (external) electric field is applied. This is similar to a free-fall motion in a gravitational field.
When a probe electric field is introduced, in response to it, additional currents will be generated from which the conductivity 
in a curved background spacetime can be obtained. When the external electric field tries to drive a current along/against the geometric
currents, the system will appear in low-/high-resistance state.

With the above motivations to study the semi-classical transport of tilted Dirac fermions, in the rest of the paper we adopt the Boltzmann equations in curved
spacetime to address the transport in such solids. Then we present our results and end the paper with discussions.

\section{Boltzmann equations in the curved spacetime}
The semiclassical dynamics of relativistic fermions in arbitrary metric background is a well studied problem~\cite{Landau1980,Poisson2004}. 
In this section we adopt the established formulation for the case of electrons in solids with tilted Dirac cone. In this case the
spacetime geometry is entirely determined by the tilt of the Dirac cone and its variation profile in the space that can be obtained from Eq.~\eqref{Metric}. 
On the solid-state side, the semi-classical dynamics of Bloch electrons is a well established approach and was 
developed in the early days of solid state physics to provide an intuitive picture of the motion of electrons in an ionic background of a lattice. 
In combination with the Boltzmann equation, many equilibrium and transport phenomena are well described~\cite{girvin2019modern,cercignani2002relativistic}. 
In the semi-classical regime, the external fields vary slowly in space and time, such that the size of the fermion wavepacket while being much larger
than the lattice spacing, is much smaller than the length scale over which the externally applied probe fields are varying. 
Therefore ${\bs p}$ and ${\bs x}$ can be simultaneously well defined and form a phase space. 
The dimensionless distribution function $f(\bs x,\bs p,t)$ in the phase space is a quantity that must be calculated.
This quantity is probability that the Bloch state with momentum $\bs p$ (in some band $n$) at time $t$ is occupied by an electron centered at $\bs x$~\footnote{
It must be noted that distribution function is a scalar quantity as long as the chemical potential lies well above
the Dirac node, thereby the inter-band processes can be ignored. At the charge neutrality point where the interband processes
become important, the distribution function must be promoted to the form $f\sigma_0+\bs g.\bs\sigma$ in the band space represented by the 
pseudospin $\bs\sigma$~\cite{Morawetz2015,Morawetz2016,Kashuba2019}}.

To enforce the underlying structure of the spacetime in the Boltzmann equation, we must formulate it within the geometric description of the relativistic system. Electrons are characterized as electron gas with rest mass $m$ (corresponding to the energy gap $2mv_F^2$), space-time coordinates $(x^{\alpha})=(v_{F} t,\bs x)$ and three-momentum vectors $(p^{\alpha})=(p^{0},\bs p)$ (in two space dimensions) where $p^0=\varepsilon/v_F$ with $\varepsilon$ being energy and $v_F$ the upper limit of speeds in the solid. Given that the length of the three momentum vectors must be equal to $mv_{F}$, the quantity $p^{0}$ is related to $\bs p$ with the relation $p^{0}=\sqrt{\vert \bs p\vert^{2}+m^{2}v_{F}^{2}}$.
The relativistic Boltzmann equation for Minkowski space-time is~\cite{cercignani2002relativistic}
\begin{equation}
p^{\alpha}\frac{\partial f}{\partial x^{\alpha}}+m\frac{\partial(fK^{\alpha})}{\partial p^{\alpha}}=Q(f,f'),\qquad\alpha=0,1,2
\label{BE_Minko}
\end{equation} 
where $Q(f,f')$ is the collision integral and $(K^{\alpha})=(K^{0},\bf F)$ is the (relativistic) force vector. The quantity $K^{0}$ is determined by the relation $K^{\alpha}p_{\alpha}=0$.

The equation of a geodesics (path of least action) in a system of arbitrary coordinates are written in terms of the three-velocity 
$U^\alpha$ as $\frac{\delta U^{\alpha}}{\delta s}=0$ where the variations is taken in a "covariant" way that accounts for non-trivial 
structure of the background spacetime~\cite{cercignani2002relativistic}. This equation simply states that the velocity is parallel transported 
along the classical path.  Using the definition of the above derivative for the three vector 
as $\frac{\delta A^{\alpha}(\lambda)}{\delta\lambda}\equiv \frac{dA^{\alpha}}{d\lambda}+\Gamma_{\nu\sigma}^{\alpha}\frac{dx^{\nu}}{d\lambda}A^{\sigma}(\lambda)$ 
the Boltzmann equation is generalized to curved space-time with the an arbitrary metric $g_{\mu\nu}[\bs\zeta(\bs r)]$
which in the absence of external forces is
\begin{equation}
p^{\alpha}\frac{\partial f}{\partial x^{\alpha}}-\Gamma_{\nu\sigma}^{i}p^{\nu}p^{\sigma}\frac{\partial f}{\partial p^{i}}=Q(f,f'),\qquad i=1,2
\label{BE_curve1}
\end{equation} 
where, $\Gamma_{\nu\sigma}^{\alpha}$ are the Christoffel connections which in the case of torsion-free spacetime
is fully determined by the space-time metric and its derivatives as~\cite{ryder2009introduction,Carroll2019}
\be
\Gamma_{\nu\sigma}^{\alpha}=\frac{g^{\alpha\beta}}{2}\left(\frac{\partial g_{\nu\beta}}{\partial x^{\sigma}}+\frac{\partial g_{\sigma\beta}}{\partial x^{\nu}}-\frac{\partial g_{\nu\sigma}}{\partial x^{\beta}}\right).
\ee
To account for the external electromagnetic force in this formulation, we use the short-hand representation of the electromagnetic field tensor $F^{\alpha\beta}=(-$\textbf{E}$,v_{F}$\textbf{B}$)$ by which we mean
\begin{equation}
F^{\alpha\beta}=
\begin{pmatrix}
0 & -E^{1} & -E^{2}\\
E^{1} & 0 & -v_{F}B^{3}\\
E^{2} & v_{F}B^{3} & 0
\end{pmatrix},
\label{EMF_Tensor}
\end{equation} 
where again in the present synthetic case the Fermi velocity $v_F$ that defines the Dirac theory in these solids replaces the speed $c$ of light in relativistic theories of
gravitation. In fact if one studies the propagation of 
transverse electric fields in such materials, it turns out that the propagation velocity is set by $v_F$ rather than $c$ ~\cite{Jalali-Mola2020}. 
The electromagnetic force for a particle with charge $q$ will be $\frac{q}{v_{F}}F^{\alpha\beta} U_{\beta}$, where $U_{\beta}=(\gamma v_{F},-\gamma\bs v)$ 
is velocity three-vector and $\gamma=1/\sqrt{1-(v/v_F)^2}$.
Finally, adding the electromagnetic force as an external force, the Boltzmann transport equation for an electronic system is written in space-time with a general metric as follows:
\begin{equation}
p^{\alpha}\frac{\partial f}{\partial x^{\alpha}}-\Gamma_{\nu\sigma}^{i}p^{\nu}p^{\sigma}\frac{\partial f}{\partial p^{i}}-\frac{e}{v_{F}}F^{\alpha\beta} p_{\beta}\frac{\partial f}{\partial p^{\alpha}}=Q(f,f'),
\label{BE_curve2}
\end{equation} 
where $q=-e$ is the electric charge of the electron. Eq.~(\ref{BE_curve2}) shows that in curved space-time, in addition to the terms related to the changes in space-time coordinates of the distribution function and the electromagnetic force, there is a connection term $\Gamma^i_{\nu\sigma}$ that arises from the geometry and depends on 
the metric and its derivatives. 

In the following, we solve Eq.~(\ref{BE_curve2}) for the one-particle distribution function at zero temperature. In the term related to electromagnetic fields, we take the magnetic field equal to zero and study the system only in the presence of an externally applied electric field. 
Since there is no applied magnetic field, we ignore the Berry phase contributions.
This can be done in the well-established approximation 
that encodes overall effect of collisions into a single time scale that encodes the scattering effects as an average scattering rate. 

\subsection{Relaxation time approximation}
The one-particle distribution function is the quantity obtained by solving the Boltzmann equation. In other words, Eq.~(\ref{BE_curve2}) is an equation of motion for one-particle distribution function. To begin solving Eq.~(\ref{BE_curve2}) for $f$, we first start with the collision integral. The exact functionality of the collision integral depends on the characteristics of the collision source and parameters such as the electron-electron interaction strength, the differential cross section of impurity scattering, and so on~\cite{girvin2019modern}. We know that equilibrium of the system requires changing the number of micro-states. Excluding small drifts due to external fields, the equilibration is not possible without collisions. The main effect of a collision term is to control the return of an initial non-equilibrium distribution to the final equilibrium distribution. However, we can consider a phenomenological model for it known as the relaxation time approximation~\cite{girvin2019modern} that summarizes the collision integral as
\begin{equation}
Q(f,f')=-\frac{m}{\tau_{r}}(f-f^{(0)}),
\label{RTA_1}
\end{equation} 
where, $\tau_{r}$ is the relaxation time, $m$ is the rest mass of particles and $f^{(0)}$ is Fermi distribution function of equilibrium at zero temperature.

To obtain a new equilibrium distribution function from the Boltzmann equation, we use the Chapman-Enskog method~\cite{chapman1990mathematical,enskog1917kinetische}. The main idea of this method is to break the equilibrium distribution function into two additive terms as $f=f^{(0)}+\Phi$. The first part is equal to the equilibrium Fermi distribution function and the second part $(\Phi)$ represents the small deviations in the Fermi distribution function, which is treated linearly in terms of the gradient of potentials. To this end, in the left side of Eq.~(\ref{BE_curve2}) the equilibrium distribution function is approximated as $f\approx f^{(0)}$. The Fermi equilibrium distribution function at temperature $T$ is $f=\frac{1}{e^{\beta(\varepsilon-\mu)}+1}$ where $\beta=\frac{1}{k_{B}T}$, the $k_{B}$ is the Boltzmann constant, $\mu(\bs  r)$ is the chemical potential and $\varepsilon=U^{\nu}p_{\nu}$ is the energy of particles. In the relativistic version of the relaxation time approximation proposed by Anderson and Whitting~\cite{anderson1974relativistic}, we use $m=\frac{U^{\nu}P_{\nu}}{v_{F}^{2}}$ to express the rest mass of particles in terms of the velocity and momentum three vectors. By substitution the relaxation time approximation in Eq.~(\ref{BE_curve2}), the Boltzmann equation becomes
\begin{equation}
p^{\alpha}\frac{\partial f^{(0)}}{\partial x^{\alpha}}-\Gamma_{\nu\sigma}^{i}p^{\nu}p^{\sigma}\frac{\partial f^{(0)}}{\partial p^{i}}-\frac{e}{v_{F}}F^{\alpha\beta} p_{\beta}\frac{\partial f^{(0)}}{\partial p^{\alpha}}=-\frac{U^{\nu}P_{\nu}}{v_{F}^{2}\tau_{r}}\Phi.
\label{BE_rta}
\end{equation} 
In the following, we discuss the second term in the left side by referring to it as geometric term to emphasize that
the geometry $\Gamma^i_{\nu\sigma}$ is some sort of background force in disguise that albeit represents the Coulomb forces
behind the space-dependent tilting of the Dirac cone. 

\subsection{\label{sec:level2}Geometric effects}
The second term in Boltzmann's transport equation (\ref{BE_rta}) arises from the structure of space-time, which incorporates the properties of curved space-time in transport through Christoffel connections. This term represents how the background geometry influnces the evolution of the occupation probability in the phase space.
Already at the single particle-level, the geometry affects the motion of particles.
To see this, we use the geodesic equation of particles in a curved space-time. Assuming a charged particle in the presence of an electromagnetic field, the geodesic equation of the particle is~\cite{ryder2009introduction}
\begin{equation}
\frac{d^{2}x^{\alpha}}{d\tau^{2}}+\Gamma_{\nu\sigma}^{\alpha}\frac{dx^{\nu}}{d\tau}\frac{dx^{\sigma}}{d\tau}=\frac{-e}{mv_{F}}F^{\alpha\beta}U_{\beta}. 
\label{Geo_rel}
\end{equation} 
The above equation can be cast into Newton-like equation by introducing
\be
\Gamma\equiv\frac{dt}{d\tau}=\frac{1}{\sqrt{-g_{00}(1+\frac{g_{0i}}{g_{00}}\frac{v^{i}}{v_{F}})^{2}-\frac{v^{2}}{v_{F}^{2}}}},
\ee 
which allows to break the acceleration $\frac{d^{2}x^{i}}{dt^{2}}=a^i$ into two terms end rewrite Eq.~\eqref{Geo_rel} as
\begin{equation}
a^i=a^i_G+a^i_E=-\frac{1}{\Gamma^{2}}\Gamma_{\nu\sigma}^{i}\frac{dx^{\nu}}{d\tau}\frac{dx^{\sigma}}{d\tau}-\frac{1}{\Gamma^{2}}\frac{e}{mv_{F}}F^{i\beta}U_{\beta},
\label{Geo_pot}
\end{equation} 
in which $v^{i}=\frac{dx^{i}}{dt}$ and $v$ is the velocity vector length and $a^i_{G/E}$ represents the acceleration arising from geometric and (external) electric forces.
Inspired by these definitions and by substitution in Eq.~(\ref{BE_rta}), the Boltzmann equation becomes
\bearr
p^{\alpha}\frac{\partial f^{(0)}}{\partial x^{\alpha}}&+m^{2}\Gamma^{2}a^i_G\frac{\partial f^{(0)}}{\partial p^{i}}+m^{2}\Gamma^{2}a^i_E \frac{\partial f^{(0)}}{\partial p^{i}}\nn\\
&-\frac{e}{v_{F}}F^{0\beta} p_{\beta}\frac{\partial f^{(0)}}{\partial p^{0}}=-\frac{U^{\nu}P_{\nu}}{v_{F}^{2}\tau_{r}}\Phi,
\label{BE_rta_gp-k}
\eearr 
where the second term on the left side of this equation directly represents the contribution of the geometric force effect and the third and fourth terms stand 
for the effect of the external electromagnetic force in the background geometry given by the metric~\eqref{Metric}. By writing the explicit form of $a^i_E$ from the definition of 
Eq. (\ref{Geo_pot}), the Boltzmann equation will be
\bearr
p^{\alpha}\frac{\partial f^{(0)}}{\partial x^{\alpha}}+m^{2}\Gamma^{2}a^i_G\frac{\partial f^{(0)}}{\partial p^{i}}-\frac{e}{v_{F}}F^{\alpha\beta} p_{\beta}\frac{\partial f^{(0)}}{\partial p^{\alpha}}=-\frac{U^{\nu}P_{\nu}}{v_{F}^{2}\tau_{r}}\Phi.\nn\\
\label{BE_rta_gp}
\eearr
Now we can solve and obtain the $\Phi$ function in terms of the external electric field, chemical potential
and geometric forces. The steady state solution ($\frac{\partial f^{(0)}}{\partial x^{0}}=0$) is
\begin{widetext}
\bearr
\Phi=&-\frac{v_{F}\tau_{r}}{\Gamma}\left( -\frac{\partial f^{(0)}}{\partial \varepsilon}\right) \left\lbrace  \frac{p^{i}}{p_{0}}\frac{\partial \mu}{\partial x^{i}}\right\rbrace
+\Gamma^{4}\tau_{r}\left( -\frac{\partial f^{(0)}}{\partial \varepsilon}\right) \left\lbrace (\zeta_{i}^{2}-g_{00})p^{i}+\zeta_{i}\zeta_{\bar{i}}p^{\bar{i}}\right\rbrace a^i_G\nn\\
&-v_{F}\tau_{r}\left( -\frac{\partial f^{(0)}}{\partial \varepsilon}\right)\left\lbrace \zeta_{i}+2(\zeta_{i}^{2}-g_{00})\frac{p^{i}}{p_{0}}+2\zeta_{i}\zeta_{\bar{i}}\frac{p^{\bar{i}}}{p_{0}}\right\rbrace \frac{\partial (-e\varphi_{E})}{\partial x^{i}}
\label{fi}
\eearr
where {$p_{0}=\sqrt{g_{00}m^{2}v_{F}^{2}+(1-\zeta_{y}^{2})(p^{1})^{2}+(1-\zeta_{x}^{2})(p^{2})^{2}+2\zeta_{x}\zeta_{y}p^{1}p^{2}}$}. 
Here $\varphi_{E}$ is the external (probe) electric potential and we have used $E^{i}=-\frac{\partial \varphi_{E}}{\partial x^{i}}$ and $\frac{\partial f^{(0)}}{\partial \mu}=-\frac{\partial f^{(0)}}{\partial \varepsilon}$ in our calculations. 
By obtaining the deviation $\Phi$ from the equilibrium distribution we can compute the current density. The Fermi equilibrium distribution function at zero temperature is a step function in which states below the Fermi energy are occupied and states above it are unoccupied. Therefore at zero temperature, derivative of Fermi distribution function with respect to energy will be a Dirac delta function at Fermi energy, $(-\frac{\partial f^{(0)}}{\partial \varepsilon})=\delta(\varepsilon-\varepsilon_{F})$. The integral of the electric current density is as follows~\cite{cercignani2002relativistic},
\begin{equation}\label{ECD_int}
J^i=(-2e)v_{F}\int p^{i}f\sqrt{g}\frac{d^{2}p}{p_{0}},
\end{equation} 
where $g=-\det(g_{\alpha\beta})=1$. By writing the function $f=f^{0}+\Phi$, the integral will have two parts, only the second part of which has a non-zero value.
Substituting $\Phi$ from Eq.~\eqref{fi} gives (see the appendix for details),
\begin{equation}
J^{i}=G^i(\zeta_{x},\zeta_{y})+\mathbb{A}\left\lbrace \left[ (1-\zeta_{i}^{2})\frac{\partial \mu}{\partial x^{i}}-\zeta_{x}\zeta_{y}\frac{\partial \mu}{\partial x^{\bar{i}}}\right]
+2\sqrt{-g_{00}}\frac{\partial (-e\varphi_{E})}{\partial x^{i}}\right\rbrace,
\label{ECD_solved}
\end{equation}
where, $G^i(\zeta_{x},\zeta_{y})=-2ev_{F}\int \frac{p^{i}}{p_{0}}\Phi_{g}d^{2}p$  in which 
$\Phi_{g}=-v^{2}_{F}\tau_{r}\left( -\frac{\partial f^{(0)}}{\partial \varepsilon}\right) \Gamma_{\nu\sigma}^{i}\frac{p^{\nu}}{p_{0}}\frac{p^{\sigma}}{p_{0}}\left[(\zeta_{i}^{2}-g_{00})p^{i}+\zeta_{i}\zeta_{\bar i}p^{\bar i}\right]$ and for $i=(1,2)$ we have $\bar i=(2,1)$ and the constant $\mathbb{A}=-2\pi e\tau_{r}\varepsilon_{F} [1+( \frac{mv_{F}^{2}}{\varepsilon_{F}})^{2}]$ is introduced. These integrals are complicated
functions of $(\zeta_x,\zeta_y)$ that can be computed both analytically and numerically with desired precision.
\end{widetext}
In the above derivation, we have used the fact that the deviation $\Phi$ of the distribution function from its equilibrium value $f^{(0)}$ is small
and is linear in terms of potentials (i.e. chemical potential, geometric forces and external electric potential). 
The above result clearly shows that the current is driven by three mechanisms: (1) the spatial variations of the chemical potential (the term in the square bracket), 
(2) the geometric forces that compactly represent the geometry of the spacetime (the term outside the curly bracket) and (3) the externally applied
electric field that has been encoded into $\varphi_E$ (the second term in the curly bracket). 
The item (2) above is the salient feature of the transport in tilted Dirac materials whose tilt varies in space. Interestingly even the item (1) above 
is highly non-trivial: The variations of the chemical potential in the lateral direction can give rise to a transverse current
\be
   J^i=\mathbb{A}\zeta_x\zeta_y\frac{\partial \mu}{\partial x^{\bar i}}
   =|\epsilon^{ij}| \mathbb{A}\zeta_x\zeta_y\frac{\partial \mu}{\partial x^j}. 
   \label{PHE.eqn}
\ee
where $\epsilon^{12}=-\epsilon^{21}=1$ and $\epsilon^{11}=\epsilon^{22}=0$. 
This term that relies on $\zeta_x\zeta_y$ is reminiscent of a similar transverse response in the
hydrodynamic study of tilted Dirac cone materials~\cite{AhmadSciPost}. Since instead of $\epsilon^{ij}$, its absolute
value enters the above equation, it appears that due to the symmetric matrix structure one can "rotate away" the 
above effect by choosing a coordinate system that lies along the principal axes. However the tilt $\zeta^i$ is part of a {\em spacetime} metric
that mingles space and time. Therefore ordinary rotations are not isometries of the new spacetime anymore~\cite{Jafari2019}, meaning that
simple rotations of the "non-tilted" spacetime around $z$-axis, do not leave the electron system invariant.
Such terms in a different context are known as planar Hall effect to distinguish them from the standard Hall effect~\cite{Zheng2020}
where the above response arises when an in-plane magnetic field is applied (hence the name "planar" Hall effect). 
The essential point is that in our case the planar Hall effect arises from spatial variations of the tilt parameter and is manifested as the 
lateral current in response to a chemical potential gradient. 
Therefore when $\zeta_{x},\zeta_y$ depend on space, Eq.~\eqref{PHE.eqn} can be considered a {\em local} version of the {\em geometric planar Hall effect}. 
To obtain the contravariant components $g^{\mu\nu}$, one needs to invert the covariant components $g_{\mu\nu}$ given in~\eqref{Metric}. 
Doing this reveals that $\zeta^x\zeta^y$ is the component $g^{12}$ or $g^{xy}$. Therefore Eq.~\eqref{PHE.eqn} provides a direct access to off-diagonal 
components of the metric and hence can be considered as a method for measuring the metric in local transport. 

As can be seen in the Boltzmann formulation the geometry of the spacetime (encoded in the Christoffel connections) enters
both at the level of equation of motion, and also the evolution of the occupation probability in the phase space.

\section{results and discussion} 
The part of the current formula that can be tuned externally can be simplified to the final form (see the appendix) 
\begin{equation}
J^{i}=\mathbb{A}\left\lbrace \left[ (1-\zeta_{i}^{2})\frac{\partial \mu}{\partial x^{i}}-\zeta_{x}\zeta_{y}\frac{\partial \mu}{\partial x^{\bar{i}}}\right]
+2\sqrt{-g_{00}}\frac{\partial (-e\varphi_{E})}{\partial x^{i}}\right\rbrace,
\label{Jmuphi.eqn}
\end{equation}
where as pointed out $\mathbb{A}=-2\pi e\tau_{r}\varepsilon_{F} [1+( \frac{mv_{F}^{2}}{\varepsilon_{F}})^{2}]$. 
In flat Minkowski space-time~\cite{girvin2019modern} the current density as in any other conductor 
is proportional to the sum of the gradients of electrical and chemical potentials giving 
$\bs J\propto\left[ {\nabla}\mu+{\nabla}(-e\varphi_{E})\right]={\nabla}\left( \mu-e\varphi_{E}\right)$. 
This qualifies the "electro-chemical" potential as the source for deriving current as the above combination of the two potentials naturally arises. 
But unlike flat space-time, the first point to be deduced from our results is that in the case of spatially varying tilt parameter, 
the electrical conductivity is not given by the joint effect of the gradient of electrical and chemical potentials with equal coefficients. 
Therefore the concept of electro-chemical potential will not be valid in solids with spatially variable tilt parameter and requires appropriate generalization. 
The coefficients for each potential gradient is determined by the tilt parameters. 
As can be seen from Eq.~\eqref{Jmuphi.eqn}, when the tilt parameters in both directions are non-zero,
the chemical potential gradient in one direction can induce a current in the transverse direction.
Such an off-diagonal conductivity is not of the Hall-type as the resulting current $J^i$ can not be expressed as the antisymmetric form $\epsilon^{ij}\partial_j\mu$. Instead
it is symmetric and proportional to $|\epsilon^{ij}|\partial_j\mu$. As pointed out since the isometries of the new spacetime~\eqref{Metric} 
are not pure rotations~\cite{Jafari2019}, the above effect can not be eliminated by rotation~\cite{AhmadSciPost} and
therefore will be a genuine geometric form of planar Hall effect. Note that the electric component $\varphi_E$ does not contribute to the planar geometric "Hall" (transverse)
conductivity. The electric potential $\varphi_E$ that gives rise to longitudinal current, encodes the red-shift factors $\sqrt{-g_{00}}$. This term will give a 
direction independent current. This feature can be used to experimentally separate the part of current that arises from $\partial_i\varphi_E$ and the part that arises from 
$\partial_i\mu$. The above quantities are local and require four probe van der Pauw measurements~\cite{vdPauw}. 

Let us return to the discussion of the first term of Eq.~\eqref{ECD_solved} where the effects of the tilting 
of the energy bands or the geometry that are present even in the absence of externally applied electric field is dumped into the term $G^i(\zeta_x,\zeta_y)$. 
In addition to the fact that the electro-chemical potential looses its meaning in spatially variable tilt situation, 
a totally new background driving force arising from the geometry of the synthetic spacetime gives rise to the terms $G^i$ in the current.
A manifestation of such term can be a "zero-biased" currents observed in the surface of topological insulators~\cite{Ogawa2016} where a background
current as well as a photocurrent can be explained in terms of tilting of the Dirac cone~\cite{Jafari2023}. 
In case of Dirac cones that are not spin-orbit coupled and hence the tilt is not easy to tune by magnetization, 
this can be considered as a materials property whose shape is fixed during the crystallization/synthesis of the material. 
Since such a term is present even when the driving external electric field absent, it can be considered the solid-state analog of gravitational
"free fall" movement of the particles in a region with non-trivial gravitational geometry.~\footnote{Of course for arbitrary geometry, the free fall path
is not necessarily straight line and in general it is determined by the geodesic equation.} 
When the driving external electric field drives the current along (opposite to) the "free fall" tendency, a given amount of electric field will result
in larger (smaller) current. Such asymmetry can be considered as a direction dependent resistivity. Extreme limit of such asymmetry can be regarded a current rectifier. 
Less extreme case may find applications by selectively allowing high/low current.
This effect can be used to detect small terahertz radiation where population of tilted Dirac band generates an associated current~\cite{Zhang2020,Rogalski2023}. 
The curved spacetime incarnation of such detectors may offer varying sensitivity depending on the local values of the tilt parameter. 

\section{Conclusions}
In conclusion, we have investigated the effect of spatially variable tilt on the semi-classical transport of 2D tilted Dirac cone materials using Boltzmann transport theory.
Such a spacetime can be realized when instead of a uniform magnetization~\cite{Ogawa2016} a planar magnetic textures is placed on the surface of a topological insulators. 
In such a case the ensuing spacetime structure gives rise to a "gravito-electric" forces that manifest as a isotropic conductivity term encoding the gravitational red-shift effects, as well as additional terms deriving from the dependence of the chemical potential $\mu$ on spatial coordinates. A unique manifestation of the gravito-electric forces will 
be a geometrical version of planar Hall effect where a current transverse to the gradient of $\mu$ arises that is determined by the off-diagonal component  $g^{xy}$ of the metric. This can be considered as a way to experimentally measure the metric by looking into local current. 

\begin{acknowledgments}
This work was initiated when SAJ was a professor at Sharif University of Technology (SUT) and completed during a visit to
Forschungszentrum, J\"ulich supported by the Alexander von Humboldt foundation.  
Hospitaility of Eva Pavarini and David DiVincenzo is appreciated.  
\end{acknowledgments}

\appendix
\section{Current Density Integral Calculation}
Here we calculate the electric and chemical potential contribution to the electric current density in more detail.
The integral of the electric current density is as follows
\begin{equation}
J^{i}=(-2e)v_{F}\int p^{i}f\sqrt{g}\frac{d^{2}p}{p_{0}},
\label{ECD_int}
\end{equation} 
where $g=-\det(g_{\alpha\beta})=1$ and for $f^{(0)}$ the integral does not have contribution. We compute the integral in the polar coordinate system and for this purpose we must consider the following relations
\begin{widetext}
\begingroup\makeatletter\def\f@size{10.5}\check@mathfonts 
\begin{equation}
\begin{aligned}
&d^2p=\vert p\vert d\vert p\vert d\theta ,\\
&p^1=\vert p\vert \cos(\theta), \quad p^2=\vert p\vert \sin(\theta) ,\\
&\varepsilon =\Gamma v_Fp_0=\Gamma v_F\sqrt{g_{00}m^2v_F^2+\vert p\vert ^2\left[1-\left(\zeta_{x}\sin(\theta)-\zeta_{y}\cos(\theta) \right)^2 \right]} ,\\
&\varepsilon = \varepsilon_{F} \Rightarrow \vert p_{F}\vert =\sqrt{\frac{\left[(\frac{\varepsilon_{F}}{\Gamma v_F})^2-g_{00}m^2v_F^2 \right]}{\left[1-\left(\zeta_{x}\sin(\theta)-\zeta_{y}\cos(\theta) \right)^2 \right] }},\\
&\left( -\frac{\partial f^{(0)}}{\partial \varepsilon}\right)=\delta(\varepsilon - \varepsilon_{F})=\frac{\delta (\vert p\vert -\vert p_{F}\vert)}{\vert \frac{\partial \varepsilon}{\partial \vert p\vert}\vert _{\vert_{\vert p_{F}\vert}}}=\frac{\varepsilon_{F}}{(\Gamma v_F)^2}\frac{\delta (\vert p\vert -\vert p_{F}\vert)}{\sqrt{\left[1-\left(\zeta_{x}\sin(\theta)-\zeta_{y}\cos(\theta) \right)^2  \right] \left[ \left(\frac{\varepsilon_F}{\Gamma v_F}\right)^2-g_{00}m^2v_F^2 \right] }}. 
\end{aligned}
\label{EC-density-rels}
\end{equation} 
\endgroup
By inserting relations of Eq. (A.2) in Eq. (A.1), the integral over the modulus of $\vec p$ (i.e. $d\vert p\vert$) can be easily calculated due to the delta function, and only three angular integrals are left as follows: 
\begingroup\makeatletter\def\f@size{10.5}\check@mathfonts 
\begin{equation}
\begin{aligned}
&(p^1)^2\Longrightarrow \int_{0}^{2\pi}\frac{\cos^2(\theta)}{\left[1-\left(\zeta_{x}\sin(\theta)-\zeta_{y}\cos(\theta) \right)^2  \right]^2} d\theta ,\\
&p^1p^2\Longrightarrow \int_{0}^{2\pi}\frac{\sin(\theta)\cos(\theta)}{\left[1-\left(\zeta_{x}\sin(\theta)-\zeta_{y}\cos(\theta) \right)^2  \right]^2} d\theta ,\\
&(p^2)^2\Longrightarrow \int_{0}^{2\pi}\frac{\sin^2(\theta)}{\left[1-\left(\zeta_{x}\sin(\theta)-\zeta_{y}\cos(\theta) \right)^2  \right]^2} d\theta
.
\end{aligned}
\label{EC-density-ints}
\end{equation} 
\endgroup
To calculate the angular integrals of Eq. (A.3), we use residue theorem. Before that, a relief comes about by using the following trick that
reduces the degree of the pole: For example the integral corresponding to $(p^1)^2$ will become
\begingroup\makeatletter\def\f@size{10.5}\check@mathfonts 
\begin{equation}
\begin{aligned}
&\int_{0}^{2\pi}\frac{\cos^2(\theta)}{\left[1-\left(\zeta_{x}\sin(\theta)-\zeta_{y}\cos(\theta) \right)^2 \right]^2} d\theta =-\frac{d}{da}\left(  \int_{0}^{2\pi}\frac{\cos^2(\theta)}{\left[a-\left(\zeta_{x}\sin(\theta)-\zeta_{y}\cos(\theta) \right)^2 \right]} d\theta\right)_{\vert _{a=1}},
\end{aligned}
\label{EC-density-ints2}
\end{equation} 
\endgroup
Now, by changing the variable as $z=e^{i\theta}$, the sine and cosine functions are written as $\sin(\theta)=\frac{-i}{2}(z-\frac{1}{z})$, $\cos(\theta)=\frac{1}{2}(z+\frac{1}{z})$ and $d\theta =\frac{dz}{iz}$. By substituting them in the integral of Eq. (A.4) we have,
\begingroup\makeatletter\def\f@size{10.5}\check@mathfonts 
\begin{equation}
\begin{aligned}
&I(a)\equiv \int_{0}^{2\pi}\frac{\cos^2(\theta)}{\left[a-\left(\zeta_{x}\sin(\theta)-\zeta_{y}\cos(\theta) \right)^2 \right]} d\theta = \oint_{\vert z\vert =1}\frac{i(z^4+2z^2+1)}{\zeta_{+}^2z^5-4(a-\frac{\zeta_{+}\zeta_{-}}{2})z^3+\zeta_{-}^2z}dz,
\end{aligned}
\label{EC-density-ints3}
\end{equation} 
\endgroup
where, $\zeta_{\pm}=\zeta_{y}\pm i\zeta_{x}$. By setting $\zeta_{+}^2 z^5 -4(a-\frac{\zeta_{+}\zeta_{-}}{2})z^3+\zeta_{-}^2 z=0$, we obtain the poles of the integrand as follows
\begingroup\makeatletter\def\f@size{10.5}\check@mathfonts 
\begin{equation}
\begin{aligned}
&z_1= \sqrt{\frac{-2a+\zeta_{x}^2+\zeta_{y}^2+2\sqrt{a^2-a(\zeta_{x}^2+\zeta_{y}^2)}}{(\zeta_{x}-i\zeta_{y})^2}} ,\\
&z_2=-\sqrt{\frac{-2a+\zeta_{x}^2+\zeta_{y}^2+2\sqrt{a^2-a(\zeta_{x}^2+\zeta_{y}^2)}}{(\zeta_{x}-i\zeta_{y})^2}} ,\\
&z_3= \sqrt{\frac{-2a+\zeta_{x}^2+\zeta_{y}^2-2\sqrt{a^2-a(\zeta_{x}^2+\zeta_{y}^2)}}{(\zeta_{x}-i\zeta_{y})^2}} ,\\
&z_4=-\sqrt{\frac{-2a+\zeta_{x}^2+\zeta_{y}^2-2\sqrt{a^2-a(\zeta_{x}^2+\zeta_{y}^2)}}{(\zeta_{x}-i\zeta_{y})^2}} ,\\
&z_5=0,
\end{aligned}
\label{EC-density-pols}
\end{equation} 
\endgroup
where, $z_1$,$z_2$ and $z_5$ are enclosed within the path and by calculating their residue, the result of the integral is obtained as
\begingroup\makeatletter\def\f@size{10.5}\check@mathfonts 
\begin{equation}
\begin{aligned}
&I(a)=2\pi i(Res(z1)+Res(z2)+Res(z5))=\\
&=\frac{2\pi\left[a^2(\zeta_{x}-\zeta_{y})^2+\zeta_{x}^2(\zeta_{x}+\zeta_{y})^2\sqrt{a^2-a(\zeta_{x}^2+\zeta_{y}^2)}-a(\zeta_{x}^4-\zeta_{y}^4)-a(\zeta_{x}^2-\zeta_{y}^2)\sqrt{a^2-a(\zeta_{x}^2+\zeta_{y}^2)} \right] }{(a^2-a(\zeta_{x}^2+\zeta_{y}^2))(\zeta_{x}^2+\zeta_{y}^2)^2}.
\end{aligned}
\label{EC-density-ints4}
\end{equation} 
\endgroup
Now, by taking the derivative ($-\frac{d}{da}$) and substituting $a=1$ for the main integral, we have
\begingroup\makeatletter\def\f@size{10.5}\check@mathfonts 
\begin{equation}
\begin{aligned}
&\int_{0}^{2\pi}\frac{\cos^2(\theta)}{\left[1-\left(\zeta_{x}\sin(\theta)-\zeta_{y}\cos(\theta) \right)^2 \right]^2} d\theta =\frac{\pi (1-\zeta_{x}^2)}{(1-\zeta_{x}^2-\zeta_{y}^2)^{\frac{3}{2}}}=\frac{\pi (\zeta_{x}^2-1)}{g_{00}\sqrt{-g_{00}}}
\end{aligned}
\label{EC-density-ints5}
\end{equation} 
\endgroup
In the same way, for other integrals, the results are obtained as follows,
\begingroup\makeatletter\def\f@size{10.5}\check@mathfonts 
\begin{equation}
\begin{aligned}
&\int_{0}^{2\pi}\frac{\sin(\theta)\cos(\theta)}{\left[1-\left(\zeta_{x}\sin(\theta)-\zeta_{y}\cos(\theta) \right)^2 \right]^2} d\theta =\frac{-\pi \zeta_{x}\zeta_{y}}{(1-\zeta_{x}^2-\zeta_{y}^2)^{\frac{3}{2}}}=\frac{\pi \zeta_{x}\zeta_{y}}{g_{00}\sqrt{-g_{00}}},\\
&\int_{0}^{2\pi}\frac{\sin^2(\theta)}{\left[1-\left(\zeta_{x}\sin(\theta)-\zeta_{y}\cos(\theta) \right)^2 \right]^2} d\theta =\frac{\pi(1-\zeta_{y}^2)}{(1-\zeta_{x}^2-\zeta_{y}^2)^{\frac{3}{2}}}=\frac{\pi (\zeta_{y}^2-1)}{g_{00}\sqrt{-g_{00}}},
\end{aligned}
\label{EC-density-ints6}
\end{equation} 
\endgroup
Substituting everything in Eq.~\eqref{ECD_int} leads to Eq.~\eqref{Jmuphi.eqn} of the main text. 
\end{widetext}

\nocite{*}

\bibliography{bibsm-b}

\end{document}